\documentclass[twocolumn,prl,showpacs]{revtex4}
\usepackage{amsmath}
\usepackage{graphicx}

\newcommand{\eq}{\begin{equation}}
\newcommand{\fine}{\end{equation}}

\begin{document}

\title{Experimental macroscopic coherence by phase-covariant cloning of a single
photon}
\author{Eleonora Nagali$^1$, Tiziano De Angelis$^1$, Fabio
Sciarrino $^{2,1}$,  and Francesco De Martini$^1$ \\
$^1$Dipartimento di Fisica dell'Universit\'a ''La Sapienza'' and Consorzio\\
Nazionale Interuniversitario per le Scienze Fisiche della Materia, Roma,
00185 Italy\\
$^2$Centro di\ Studi e Ricerche ''Enrico Fermi'', Via Panisperna
89/A,Compendio del Viminale, Roma 00184, Italy}

\begin{abstract}
We investigate the multiphoton states generated by high-gain optical
parametric amplification of a single injected photon, polarization encoded
as a ''qubit''. The experiment configuration exploits the optimal
phase-covariant cloning in the high gain regime. The interference fringe
pattern showing the non local transfer of coherence between the injected
qubit and the mesoscopic amplified output field involving up to 4000 photons
has been investigated. A probabilistic new method to extract full
information about the multiparticle output wavefunction has been implemented.

\end{abstract}

\maketitle

The problem of manipulating and controlling the flux of quantum information
from a quantum system to many ones has been generally tackled and solved by
the theory of quantum cloning \cite{Scar05,Cerf05,DeMa05b}. Indeed, from a
practical point of view, this process is considered a useful tool for the
optimal distribution of quantum information carried by $N$ qubits into
several quantum channels and for the transmission of information contained
in a system into correlations between many systems. On a more fundamental
level, it has been suggested that, in the domain of quantum optics, the
cloning process can also be adopted to investigate the intriguing transition
of the quantum behavior from the ''\textit{microscopic world}'', represented
here by one of few properly encoded photons, to the mesoscopic and
macroscopic one composed by many particles \cite{Zure03,DeMa98}.

In quantum optics, the qubit is generally implemented by exploiting the
polarization state $\overrightarrow{\pi }$ of a single photon. It is there
natural to associate a cloning map with the process of stimulated emission
\cite{Simo00}. In the present framework, the quantum injected - optical
parametric amplifier (QI-OPA), which exploits the process of stimulated
emission in a nonlinear (NL) crystal, has led to the experimental
realization of the $1$ to $2$ cloning machine in different configurations
\cite{DeMa02,Lama02,Pell03}. Recently, the coherence-preserving character of
high-gain OPA allowed the reproduction of the quantum superposition
condition of an input single-particle qubit $\alpha \left| H\right\rangle
+\beta \left| V\right\rangle $, where $\left| H\right\rangle $ and $\left|
V\right\rangle $ are respectively single photon state of horizontal ($%
\overrightarrow{\pi }_{H}$) and vertical ($\overrightarrow{\pi }_{V}$)
polarization, in a multi-particle amplified output field involving an
average of $\simeq $ $6$ clones \cite{DeMa05}. The adoption of more powerful
amplifier operating at the quantum level could open new perspectives for the
generation of unique quantum states for the implementation of new tests of
fundamental physics and innovative protocols for quantum metrology \cite
{Agar06}.

\begin{figure}[h]
\includegraphics[scale=.3]{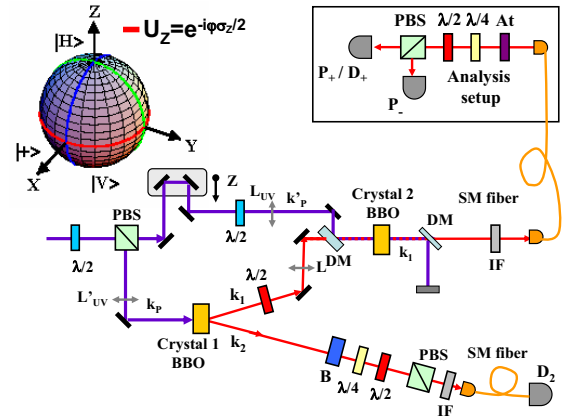}
\caption{Optical configuration of the quantum injected optical parametric
amplifier. The SPDC quantum injector (crystal 1) is provided by a type II
generator of polarization entangled photon couples. Crystal 2, realizing the
OPA action, is cut for collinear type II phase matching. Both crystals are
1.5 mm thick. The fields are coupled to single mode (SM) fiber. The UV beam
was focused on crystal 2 with $L^{\prime}_{UV}$ (focal length $= 30$ cm).
Bloch sphere representation of the input qubit.}
\end{figure}

The present work reports the optical parametric amplification of a single
photon in the high gain regime to experimentally investigate how the
information initially contained in its polarization state is distributed
over a large number of particles. In particular we analyze how the coherence
properties of the input state are transferred to the mesoscopic output
field. Fringe patterns have been observed for output fields involving up to $%
4\times 10^{3}$ photons.

The adopted QI-OPA configuration is the highly efficient collinear regime
\cite{DeMa98B}, which generates pairs of photons over the same output
spatial mode $\mathbf{k}_{1}$. This transformation realizes the optimal
distribution of quantum information for qubits restricted to a subspace of
the polarization space \cite{Scia05}. Specifically, here we consider the
optimal process for equatorial qubit of the corresponding Bloch sphere: $%
\left| \varphi \right\rangle =2^{-1/2}(\left| H\right\rangle +e^{i\varphi
}\left| V\right\rangle )$ (Fig.1); since the information is codified in the
phase $\varphi $ of the input qubit, the process is called phase-covariant
cloning \cite{Brus00,DAri03}. The partial a-priori information on the qubit
to be cloned leads to a high visibility coherence of the generated
multiphoton superposition.

The experimental layout is sketched in Figure 1. The excitation source is a
Ti:Sa Coherent MIRA mode-locked laser further amplified by a Ti:Sa
regenerative REGA device operating with pulse duration $180fs$ at a
repetition rate of $100kHz$ with average output power equal to $720$ $mW$.
The output beam, frequency-doubled by second harmonic generation, provides
the excitation beam of UV wavelength (wl) $\lambda _{P}=397.5nm$ and power $%
300mW$. The UV beam is splitted in two beams through a $\lambda /2$
waveplate and a polarizing beam splitter (PBS) and excites two BBO ($\beta $%
-barium borate) NL crystals cut for type II phase-matching. Crystal 1,
excited by the beam $\mathbf{k}_{P}$, is the spontaneous parametric
down-conversion (SPDC) source of entangled photon couples of wl $\lambda
=2\lambda _{P}$, emitted over the two output modes $\mathbf{k}_{i}$ ($i=1,2$%
) in the \textit{singlet} state $\left| \Psi ^{-}\right\rangle _{k1,k2}$=$%
2^{-%
{\frac12}%
}\left( \left| H\right\rangle _{k1}\left| V\right\rangle _{k2}-\left|
V\right\rangle _{k1}\left| H\right\rangle _{k2}\right) \label{SPDCentangled}$%
. The pump power of beam $\mathbf{k}_{P}$ is set in order to have a
negligible probability to generate two pairs of photons. While the photon
associated to mode $\mathbf{k}_{2}$ is coupled into a single mode fiber and
excites the single photon counting module (SPCM) $D_{2}$ (hereafter referred
to as \textit{trigger} mode), the single photon state generated over the
mode $\mathbf{k}_{1}$ is injected, together with an UV pump beam (mode $%
\mathbf{k}_{P}^{\prime }$), into the non-linear crystal 2 and stimulates the
emission of many photon pairs. By virtue of the nonlocal correlation acting
on modes $\mathbf{k}_{1}$ and $\mathbf{k}_{2}$, the input qubit on mode $%
\mathbf{k}_{1}$ is prepared in the state $\left| \varphi \right\rangle
_{k1}=2^{-%
{\frac12}%
}(\left| H\right\rangle _{k1}+e^{i\varphi }\left| V\right\rangle _{k1})$ by
measuring the photon on mode $\mathbf{k}_{2}$ in the appropriate
polarization basis. This measurement is carried out by adopting a set $%
\lambda /2$ + $\lambda /4$ waveplates, a phase shifter ($B$) and a
polarizing beam-splitter $PBS$. By an adjustable spatial delay ($Z$), the
time superposition in the OPA of the excitation UV pulse and of the
injection photon wavepacket is ensured. The injected single photon and the
UV pump beam $\mathbf{k}_{P}^{\prime }$ are superposed exploiting a dichroic
mirror ($DM$) with high reflectivity at $\lambda_{P}$ and high
transmittivity at $\lambda $.

The crystal 2 is oriented for collinear operation over the two linear
polarization modes, respectively horizontal and vertical. The interaction
Hamiltonian of the optical parametric amplification (crystal 2) $\widehat{H}%
=i\chi \hbar \left( \widehat{a}_{H}^{\dagger }\widehat{a}_{V}^{\dagger
}\right) +h.c.$ acts on the single spatial mode $\mathbf{k}_{1}$ where $%
\widehat{a}_{\pi }^{\dagger }$ is the one photon creation operator
associated to $\mathbf{k}_{1}$ mode with a polarization $\overrightarrow{\pi
}$. The main feature of this Hamiltonian is its peculiar property of
phase-covariance for qubits with equatorial polarization, leading to the
optimality of the cloning process. Owing to this invariance under $U(1)$
transformations, we can then re-write: $\widehat{H}$ =$\frac{1}{2}i\chi
\hbar e^{-i\varphi }\left( \widehat{a}_{\varphi }^{\dagger 2}-e^{i2\varphi }%
\widehat{a}_{\varphi \perp }^{\dagger 2}\right) +h.c.$ for $\varphi \in
(0,2\pi )$ where $\widehat{a}_{\varphi }^{\dagger }=2^{-1/2}(\widehat{a}%
_{H}^{\dagger }+e^{i\varphi }\widehat{a}_{V}^{\dagger })$ and $\widehat{a}%
_{\varphi \perp }^{\dagger }=2^{-1/2}(-e^{-i\varphi }\widehat{a}%
_{H}^{\dagger }+\widehat{a}_{V}^{\dagger })$. Any injected state $\left|
\varphi \right\rangle _{k1}$ on mode $\mathbf{k}_{1}$evolves into the output
state $\left| \Phi \right\rangle _{k1}^{\psi }=\widehat{U}\left| \psi
\right\rangle _{k1}$ according to the OPA\ unitary $\widehat{U}=\exp \left[
-i\widehat{H}t/\hbar \right] $, being $t$ the interaction time \cite{Pell03}%
. In particular we shall consider the action over states with polarization $%
\overrightarrow{\pi }_{\pm }=2^{-%
{\frac12}%
}(\overrightarrow{\pi }_{H}\pm \overrightarrow{\pi }_{V})$. The overall
output state amplified by the OPA apparatus is found to be expressed as:
\begin{equation}
\begin{aligned} &\left| \Sigma \right\rangle _{k1,k2}=(\widehat{U}\otimes
\widehat{1})\left| \Psi ^{-}\right\rangle _{k1,k2}= \\&=2^{-1/2}\left(
\left| \Phi \right\rangle _{k1}^{+}\left| -\right\rangle _{k2}-\left| \Phi
\right\rangle _{k1}^{-}\left| +\right\rangle _{k2}\right)
\label{outputstate} \end{aligned}
\end{equation}
with
\begin{equation}
\left| \Phi \right\rangle ^{\pm }=\frac{\left| \Phi \right\rangle ^{H}\pm
\left| \Phi \right\rangle ^{V}}{\sqrt{2}}=\sum\limits_{i,j=0}^{\infty
}\gamma _{ij}\frac{\sqrt{(1+2i)!(2j)!}}{i!j!}\left| 2i+1\right\rangle _{\pm
}\left| 2j\right\rangle _{\mp }
\end{equation}
and $\gamma _{ij}\equiv C^{-2}(-\frac{\Gamma }{2})^{i}\frac{\Gamma }{2}^{j}$%
, $C\equiv \cosh g$, $\Gamma \equiv \tanh g$, being $g$\ the NL\ gain\emph{\
} \cite{DeMa02}. There $\left| p\right\rangle _{+}\left| q\right\rangle _{-}$
stands for a state with $p$ photons with polarization $\overrightarrow{\pi }%
_{+}$ and $q$ photons with $\overrightarrow{\pi }_{-}.$ The multi-particle
states $\left| \Phi \right\rangle ^{+}$, $\left| \Phi \right\rangle ^{-}$are
orthonormal, i.e.$^{i}\left\langle \Phi |\Phi \right\rangle ^{j}=\delta
_{ij} $. Similar expressions holds for any equatorial pair of polarizations $%
\left\{ \overrightarrow{\pi }_{\varphi },\overrightarrow{\pi }_{\varphi
\perp }\right\} $ due to the phase-covariance of the overall process: $%
\left| \Phi \right\rangle ^{\varphi }=2^{-1/2}(\left| \Phi \right\rangle
^{H}+e^{i\varphi }\left| \Phi \right\rangle ^{V})$. Note that the state $%
\left| \Sigma \right\rangle _{k1,k2}$,\ generally dubbed ''Schroedinger Cat
State'' keeps its \textit{singlet} character in the multi-particle regime,
and expresses the nonlocal correlations between two distant objects: the
\textit{microscopic (}sigle particle) system expressed by the \textit{trigger%
} state (mode $\mathbf{k}_{2}$) and the \textit{mesoscopic (}multiparticle)
system ($\mathbf{k}_{1}$) \cite{Sch35,Schl01}.

Let us analyze the output field $\mathbf{k}_{1}$ over the polarization modes
$\overrightarrow{\pi }_{\pm }$ when the state $\left| \varphi \right\rangle
_{k1}$ is injected. The average photon number $N_{\pm }$ over $\mathbf{k}%
_{1} $ with $\overrightarrow{\pi }_{\pm }$ is found to depend on the phase $%
\varphi $ as follows: $N_{\pm }(\varphi )=\overline{m}+\frac{1}{2}(2%
\overline{m}+1)(1\pm \cos \varphi )$ with $\overline{m}=\sinh ^{2}g$. The
conditions $\varphi =0$ and $\varphi =\pi $ correspond to single-photon
injection and no-injection of the mode $\overrightarrow{\pi }_{+}$,
respectively. The average photon number related to both cases is: $N_{+}(0)=3%
\overline{m}+1$ and $N_{+}(\pi )=\overline{m}$.\ For $\varphi =0$, the
average number of photons emitted over the two polarizations over $\mathbf{k}%
_{1}$ is found to be $M=4\overline{m}+1$. We conclude that the output state
with polarization $\overrightarrow{\pi }_{\pm }$ exhibits a sinusoidal
fringe pattern of the field intensity depending on $\varphi $ with a
gain-dependent visibility $\mathcal{V}^{th}=\frac{2\overline{m}+1}{4%
\overline{m}+1}$ \cite{DeMa98B}. Note that for $g\rightarrow \infty $, viz. $%
M\rightarrow \infty $ the fringe visibility of this $1^{st}-order$
correlation function attains the asymptotic values $\mathcal{V}^{th}=50\%$.

The output field emerging from the crystal $2$ with wl $\lambda $ was
filtered against the UV pump beam by a dichroic mirror and by an
interferential filter ($IF$) with bandwidth equal to $1.5nm$. It was then
coupled to a single mode fiber, polarization analyzed and then detected by
the photomultipliers tubes ($P_{+}$ and $P_{-}$). These were Burle A02
10-dynode detectors with a Ga-As photocathode having $\eta _{QE}=13\%$. The
signals were registered by a fast digital oscilloscope [Tektronix TDS5104D]
triggered by $D_{2}$.

In a first experiment the gain value $g$ of the optical parametric process
and the overall quantum efficiency of the detection apparatus were measured
in absence of quantum injection. We obtained $g_{\max }=\left( 4.34\pm
0.02\right) $, while the overall detection efficiency $\eta _{1}$ on the $%
\mathbf{k}_{1}$ mode was estimated $\eta _{1}\simeq 1.6\%$. The latter one
was contributed by the fiber coupling ($\sim 50\%$), $IF$ transmittivity ($%
\sim 25\%$) and detector's quantum efficiency ($\eta _{QE}$). By setting $%
g=0 $ the effective visibility of the input qubit was measured: $\mathcal{V}%
_{in}=\left( 78.4\pm 5.1\right) \%$ , a result attributed to double-pairs
emission from crystal 1. By subtracting these accidental coincidences the
input qubit visibility was raised to $\left( 96.5\pm 2.5\right) \%$.

\begin{figure}[t]
\includegraphics[scale=.4]{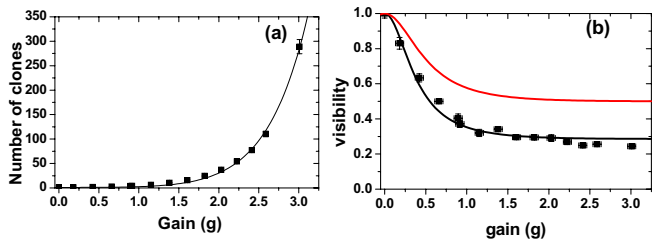}
\caption{(a) Number of clones versus the non-linear gain. Black curve: best
fit; (b) Visibility of the fringe patterns versus the gain value rescaled by
$\mathcal{V}^{in}$. Solid line : $\widetilde{\mathcal{V}}^{th}$ with and $%
p=0.4$; dashed line: $\mathcal{V}^{nc}$; dotted line $\mathcal{V}^{th}$.}
\label{fig4}
\end{figure}

In order to investigate how the coherence of the input qubit survives to the
amplification process, we measured the fringe pattern visibilities of the
output field for different values of the gain value of the OPA. The
interference property of the output field implied by the quantum
superposition character of the input qubit $\left| \varphi \right\rangle
_{k1}$ was measured in the basis $\overrightarrow{\pi }_{\pm }$ over the
output ''cloning'' mode $\mathbf{k}_{1}$ by signals of $P_{+}$ and $P_{-}$
conditioned by the detection of a single photon from $D_{2}$ for different
phases of the injected qubit. Figure 2 reports the gain dependence of the
number of clones rescaled by the visibility of the input qubit $\mathcal{V}%
_{in}$. The solid line reports the best fit curve with a theoretical model $%
\widetilde{\mathcal{V}}^{th}$ which incorporates the main experimental
imperfections: $\widetilde{\mathcal{V}}^{th}=\frac{p(2\overline{m}+1)}{p(2%
\overline{m}+1)+2\overline{m}}$. There the parameter $p$ represents the
probability that the injected heralded photon over the detected modes gives
rise to stimulated emission. It differs from $1$ because of the partial
mismatches between the amplifying pump beam and the input single photon
state. By fitting the experimental data we found $p$=$0.40\pm 0.01$
(Fig.3-b), which is in agreement with a theoretical estimation obtained by
the expression $p\simeq T\times \Delta \nu \times \Delta k$ where $T$ is
transmittivity between the two crystals ($T=0.90$), $\Delta \nu $ is the
spectral matching ($\Delta \nu \sim 0.8$) and $\Delta k$ is the spatial
matching ($\Delta k\sim 0.6$). To enlighten the resilience of coherence of
the output state, we can compare in Figure 2b the experimental data with the
dotted curve which refers to $\mathcal{V}^{th}$ and the dashed curve $%
\mathcal{V}^{nc}$ which represents the visibility that would be observed in
the case of no preservation of the coherence during the amplification ($%
\mathcal{V}^{nc}=(4\overline{m}+1)^{-1}$).

\begin{figure}[h]
\includegraphics[scale=.2]{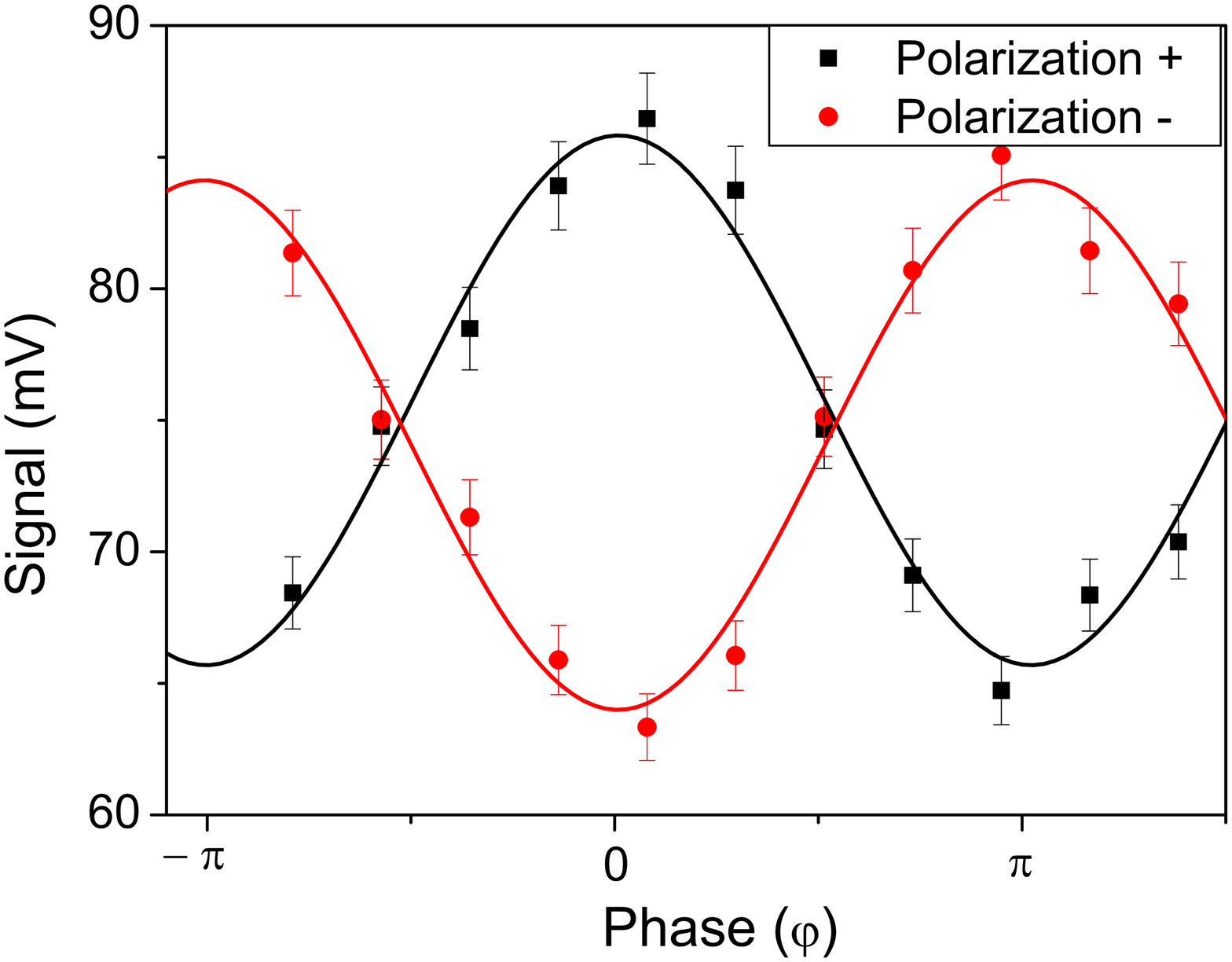}
\caption{ Average signal versus the phase of the input qubit.}
\end{figure}

The direct correlation measurement was also carried out for a gain value
equal to $g_{max}$: Figure 3. There the corresponding signals $I_{+}$ and $%
I_{-}$, averaged over $2500$ triggered pulse, are reported versus the phase $%
\varphi $ of the input qubit. The overall number of photons is equal to $%
M\sim 4000$ while the rescaled visibility of the fringe pattern is found: $%
\mathcal{V}=$ $(19.4\pm 1.8)\%$. These coherence patterns are the main
experimental result of the present article as it shows how a\emph{\ single
photon quantum superposition state} coherently drives a field with large
number of photons. Very important, the interference fringe patterns given in
Figure 3 have been experimentally observed for all orthogonal bases
belonging to the equatorial plane of the Bloch sphere of Fig.1 e.g., $\{%
\overrightarrow{\pi }_{R}=2^{-%
{\frac12}%
}(\overrightarrow{\pi }_{H}-i\overrightarrow{\pi }_{V}),\overrightarrow{\pi }%
_{L}=\overrightarrow{\pi }_{R}^{\perp }\}$ confirming the coherence between $%
\left| \Phi ^{H}\right\rangle $ and $\left| \Phi ^{V}\right\rangle $.

\begin{figure}[t]
\includegraphics[scale=.45]{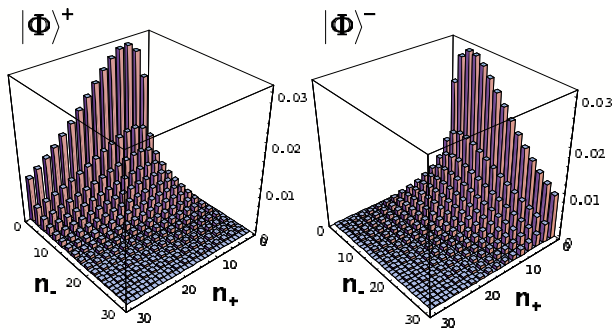}
\caption{Theoretical distribution probability of $(n_{+},n_{-})$ for $%
|\Phi\rangle^{\pm}$ $(g=1.6)$.}
\label{CalibrazFoto}
\end{figure}

We address now the problem of the distinguishability between $\left| \Phi
\right\rangle ^{\varphi}$ and $\left| \Phi \right\rangle ^{\varphi\bot}$,
which are mutually orthogonal as shown by Equation 2, e.g., $\left| \Phi
\right\rangle ^{\pm}$. A perfect discrimination between these two states can
be achieved in principle in a single run experiment by checking whether the
number of photons over the $\mathbf{k}_{1}$ mode with polarization $%
\overrightarrow{\pi }_{+}$ is even or odd. Such measurement corresponds to
estimating the operator $s_{z}=\sum_{n=0}^{\infty }\left[ \left|
2n+1\right\rangle \left\langle 2n+1\right| -\left| 2n\right\rangle
\left\langle 2n\right| \right] $ over the field $k_{1}$ with polarization $%
\overrightarrow{\pi }_{+}$. Indeed is: $s_{z}^{1+}\left| \Phi ^{\pm
}\right\rangle =\pm \left| \Phi ^{\pm }\right\rangle $. The exact
implementation of the operator $s_{z}^{1+}$ requires a photon-number
resolving detection of a mesoscopic field $\mathbf{k}_{1}$, i.e. made up by
thousands of photons. This is indeed well beyond the range of possibilities
offered by the present technology \cite{Port06}. Hence we could ask whether
the need for such extremely fine-grained resolution could be bypassed by
adopting a different approach. Let us report in Figure 4 the 3-dimensional
representations of the two probability distributions of the number of
photons which are associated with the two states $\left| \Phi \right\rangle
^{\pm }$. \ These distributions are drawn as functions of the variables $m$
and $n$ which are proportional to the size of the electronic signals $I_{+}$
and $I_{-}\;$simul$\tan $eously registered, in any single-shot experiment,
by the detectors $P_{+}$ and $P_{-}$ , respectively. Precisely, $%
I_{+}\propto m\simeq 2i$, $I_{-}\propto n\simeq 2j$, according to Equation
2. Let us refer to $\prod^{\pm }(m,n)$ as the probability that a single-shot
measurement $(I_{+},I_{-})\;$= $(m,n)$ is actually due to a realization of
the state$\;\left| \Phi \right\rangle ^{\pm }$. Since by Eq. 2 is found $%
\prod^{+}(m,n)/\prod^{-}(m,n)$ = $m/n\equiv h\;$ for $2i>>1$ and $2j>>1$, we
may conclude that the measured signal $(I_{+},I_{-})$ can be attributed to
the realization of the state $\left| \Phi \right\rangle ^{+}$ or to $\left|
\Phi \right\rangle ^{-}\;$if the measured quantity $h=I_{+}/I_{-}=m/n$ is $%
h>>1$ or $h<<1$ , respectively. In this way we can exploit the \emph{%
macroscopic character} of the output field, a continuous variable system, by
a simple photon detection technique. \ By finding an appropriate \textit{%
filtering function} $W(m,n)$, viz a weighting function of the output data $%
(I_{+},I_{-})$, the visibility of the detected fringe pattern can be raised
to a value close to $1$.

\begin{figure}[h]
\includegraphics[scale=.2]{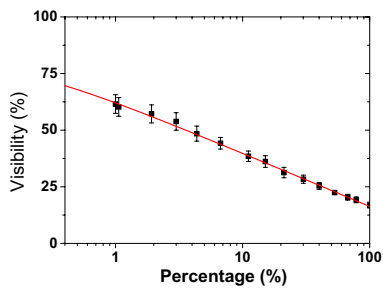}
\caption{ Visibility of the intensity versus the percentage of data which
has not been discarded within the filtering procedure. The solid line
reports the best fit with the function $f(x)= a - b \ln (x+c)$. The
percentage of discarded data scales $\simeq e^{-q}$.}
\label{visgn}
\end{figure}

Let us now implement experimentally the \textit{filtering technique}. In
order to do that in a sensible way we should consider two sources of errors:
(1) The limited detection efficiency $\eta <<1$. However this effect may
smooth the probability distributions $\prod^{\pm }(m,n)$ with no significant
effect on the ratio $h$. (2) The spontaneous emission contribution added to
the ''stimulated'' one. A discrimination between these latter contributions
is obtained by selecting output data with large $\left( I_{+}+I_{-}\right) $%
, which are most likely generated in the stimulated regime. In our
single-shot experiment we discarded the data with $|m-n|<q$ for different
values of $q$. Of course, since any such selection reduces the rate of
measurement, a convenient trading off was found necessary. Figure 5 reports
the visibilities of the fringe patterns obtained by filtering the
experimental data related to Figure 3 versus the percentage of registered
data. The behavior is found in agreement with a refined theoretical
filtering model. Interestingly, by the visibility values it is possible to
demonstrate the quantum character of the mesoscopic output field. Indeed if
the coherence of a general input qubit was not preserved during the
amplification, the process could be described as a classical channel.
Consequently the fidelity of the overall transformation would be, at most,
equal to the optimal estimation fidelity for equatorial qubits: $F_{est}=%
\frac{3}{4}$ \cite{Brus00,note}. Fig.5 achieves fidelity values above the
classical threshold value: $F=\frac{(1+V)}{2}>F_{est}$ confirming the
quantum character of the process. The same visibility have been observed for
different pairs of input equatorial qubit, as said. The ability to
discriminate the output $\{\left| \Phi \right\rangle ^{\varphi },\left| \Phi
\right\rangle ^{\varphi \bot }\}$by the above technique opens interesting
perspectives for testing quantum nonlocality in the ''macroscopic'' regime,
such as a verification of the Bell's theorem.

In summary, we have demonstrated the interference properties of the output
state in agreement with the quantum theoretical results. We reported the
experimental realization of a quantum cloning machine within two new
scenarios: (a) the optimal phase-covariant cloning based on the process of
stimulated emission, (b) the very large number of clones $M=4000>>1$. The
present work represents the first step toward an experimental interface
between single qubit systems and continuous variable ones. The attempt to
shed light on the quantum-to-classical transition by distributing the
information from a single quantum system to many ones deserves further
investigation. This work was supported by the PRIN 2005 of MIUR and project
INNESCO of CNISM.

\emph{Note added in proof}- Owing to experimental improvements of
the pumping laser power, the following parameters have been
attained: $g\simeq 5.7$, $\overline{m} \simeq 23000$.

\end{document}